\documentclass[conference]{IEEEtran}
\usepackage{epsf,amsfonts,amssymb,times,cite,amsmath,epsfig}
\usepackage{flushend}
\usepackage{graphicx}
\usepackage{epic,eepic}
\usepackage{dcolumn}
\usepackage{bm}
\def\C{{\mathbb C}}

\let\q=\quad
\let\arr=\rightarrow
\let\co=\it 
\def\CS{{\mathbb C}}
\def\im{\textrm{Im}}
\def\qq{\qquad}
\def\vq{\,,\qq}
\def\vl{\;,\;}
\def\rank{\mathop{\rm rank}\nolimits}
\def\ET{\smallbreak\par\noindent}
\def\ev#1{\medbreak\noindent{\bf#1.\enspace}}

\def\fract#1#2{{\textstyle\frac{#1}{#2}}}
\def\C(#1){{\cal #1}}

\ifCLASSINFOpdf
\else
\fi
\begin{document}
%
\title{Helstrom's Theory on Quantum Binary Decision Revisited}

\author{\IEEEauthorblockN{Gianfranco Cariolaro and Alberto Vigato}
\IEEEauthorblockA{Department of Information Engineering (DEI), University of Padova, via G. Gradenigo 6/B - Padova, PD I--35131, Italy\\
Email: \{cariolar,alvigato\}@dei.unipd.it}
}


%


\maketitle

\begin{abstract}
For a binary system specified by the density operators
$\rho_0$ and $\rho_1$ and by the prior probabilities $q_0$ and $q_1$,
Helstrom's theory permits the evaluation of the optimal measurement
operators and of the corresponding maximum  correct detection probability.
The theory is based on the eigendecomposition (EID) of the operator,
given by the difference of the weighted density operators, namely
$D=q_1\rho_1 - q_0\rho_0$. In general,  this EID is obtained explicitly
only with pure states, whereas with mixed states it must be carried out
numerically. In this letter we show that the same  evaluation
can be performed on the basis of a modified version of the Gram matrix.
The advantage is due to the fact that the outer products of density
operators are replaced by inner product, with a considerable
dimensionality reduction. At the limit, in quantum optical communications the density
operators have infinite dimensions, whereas the inner products are simply
scalar quantities.
The Gram matrix approach permits the explicit (not numerical)
evaluation of a binary system performance in cases not previously developed.
\end{abstract}


%
\IEEEpeerreviewmaketitle

\section{\label{sec:Intro}Introduction}

In 1970 Helstrom {\it et. al.} \cite{Helstrom} evaluated the optimal 
performance (minimum error probability) of a binary quantum  system, both with
pure and mixed states. Till now, this result represents the only explicit
 solution of quantum optimization. In fact, for multilevel quantum systems
the optimal solution is known in very few particular cases.

The formulation of Helstrom's  theory is the following. Let
be $\rho_0$ and $\rho_1$ the density operators of  the given 
binary quantum system  and let  $q_0$ and $q_1$ be the corresponding
prior probabilities. Then, form the {\it decision operator} as
$D=q_1\rho_1 - q_0\rho_0$ and consider its eigendecompositon (EID),
which has the general form
\begin{equation}
 D=q_1\rho_1 - q_0\rho_0=\sum_{k}{\eta_k |\eta_k\rangle
\langle \eta_k |} 
\label{eq:DD6}
\end{equation}
where $\eta_k$ are the eigenvalues and $|\eta_k\rangle$ the corresponding
eigenvectors. Finally, the correct detection probability is given by
\begin{equation}
P_c=q_0+\sum_{\eta_k>0}{\eta_k}
\label{eq:DD12}
\end{equation}
and the optimal measurement operators  by
\begin{equation}
     \Pi_1=\sum_{\eta_k>0}{|\eta_k\rangle \langle \eta_k |}\vq
     \Pi_0=\sum_{\eta_k<0}{|\eta_k\rangle \langle \eta_k |}\,.
\label{eq:DD13}
\end{equation}
 However, the explicit evaluation poses severe difficulties when the dimensions
 of the underlying Hilbert space is larger than two \cite{Bergou}.

Here, we reconsider Helstrom's theory presenting a method based on
 a ``modified'' version of the Gram matrix, which we call {\it skew
 Gram matrix}. The net difference is  that,
while the operator $D$ is expressed in terms of {\it outer products}
(see (\ref{eq:DD6})) and has the dimension of the Hilbert space,
 the Gram matrix is defined in terms of {\it inner products}.
The advantage of the replacement of  outer products 
 by inner product is a considerable
dimensionality reduction. At the limit, in quantum optical communications 
the density operators have infinite dimensions, as stated by Glauber's theory on
 the representation of quantum laser radiation field
 \cite{Glauber}, whereas the inner products are  simply
scalar quantities. 

 This new approach allows to find explicit results for the optimal
 detection, not only for the case
 of pure states (rank 1 density operators), as found in \cite{Helstrom, Helstrom2}, but also with
 mixed states. The explicit results  are particularly simple 
  when the quantum states have the {\it geometrical
 uniform symmetry} (GUS).

\section{Binary detection based on the skew Gram matrix}

Let $k_i=\rank(\rho_i)$ and assume that $k_0+k_1\leq n$,
where $n$ is the dimension of the underlying Hilbert space. 
Let $\gamma_i$ be the $n\times k_i$ factors of $\rho_i'=q_i\rho_i$,
so that
$ \gamma_0\gamma_0^*=\rho_0'$ and $ \gamma_1\gamma_1^*=\rho_1' $ ($^*$ denotes the conjugate transpose).
Then, the state matrix is given by
$
\Gamma=[ \gamma_0,\gamma_1]
$
and the Gram matrix by
\begin{equation}
 G= \Gamma^*\Gamma=\left[\begin{array}{cc}
                    \gamma_0^*\gamma_0 & \gamma_0^*\gamma_1 \\
                    \gamma_1^*\gamma_0&\gamma_1^*\gamma_1
                   \end{array}\right]=
	\left[\begin{array}{cc}G_{00}&G_{01}\\
	         G_{10}&G_{11}
 \end{array}\right]
\end{equation}
where the blocks $G_{ij}=\gamma_i^*\gamma_j$ has dimension $k_i\times k_j$.
The  skew Gram matrix (SGM) is obtained by changing the sign of the first block row,
namely
\begin{equation}
 G_s=	\left[\begin{array}{rr}-G_{00}&-G_{01}\\
	         G_{10}&G_{11}
\end{array}\right]\,.
\label{eq:DD10}
\end{equation}

Note that the ordinary Gram matrix $G$ is Hermitian and positive semidefinite (PSD), and hence $G_{00}$, 
$G_{11}$ are Hermitian and PSD, while $G_{10}=G_{01}^*$. 
The SGM is neither Hermitian nor PSD.
Note also that, given the density operator $\rho_i'$, the factors $\gamma_i$ are not unique, but the consequence of their multiplicity is irrelevant for quantum detection \cite{Elder}.

Let us first recall an importat statement from matrix analysis


\vspace{1mm}
\noindent\textsc{\textbf{Lemma 1}}	{\em \ 
 For each pair of complex matrices $A\in\CS^{m\times n}$ and $B\in\CS^{n\times m}$ with $m\leq n$, the square matrix $BA\in\CS^{n\times n}$ has the same eigenvalues as $AB\in\CS^{m\times m}$, counting both algebraic and geometrical multiplicity, together with an additional $n-m$ eigenvalues equal to 0.}
\vspace{1mm}

For the proof of this Lemma, we remand to \cite{Horn_book85}.
Now, we prove the following 


\vspace{1mm}
\noindent\textsc{\textbf{Theorem 1}}	{\em \ The $k_0+k_1$ non--zero eigenvalues $\eta_k$ of the decision operator $D=q_1\rho_1 - q_0\rho_0$ can be evaluated as the eigenvalues of the SGM $G_s$ defined by (\ref{eq:DD10}).
Moreover, the related eigenvectors $|\eta_k\rangle$ of $D$ then the optimum measurement operators
from (\ref{eq:DD13}), can be evaluated from the eigenvectors $\theta_k$ of the SGM for the same eigenvalue $\eta_k$ as 
$
|\eta_k\rangle=c_k\Gamma \theta_k
$
where the factor $c_k=1/\sqrt{\theta_k^*G\theta_k}$ assures $\langle\eta_k|\eta_k\rangle=1$.
}
\vspace{1mm}


\smallskip
In words, the eigenvalues required to evaluate the correct detection probability
 according to (\ref{eq:DD12}), can be obtained 
by solving the eigenvalue equation
\begin{equation}
 \det\left[\begin{array}{cc}-G_{00}-\eta I_{k_0}&-G_{01}\\
	         G_{10}&G_{11}-\eta I_{k_1}\end{array}\right]=0
\label{eq:D14}
\end{equation}
where $I_k$ denotes the identity matrix of order $k$.

\ET{\textbf{Proof:}} Let
$
\tilde{I} = \left[\begin{smallmatrix} -I_{k_0} & 0 \\ 0 & I_{k_1} \end{smallmatrix}\right].
$
By defining $\Delta=\tilde{I}\Gamma^*$, it is easy to prove that $D=\Gamma\Delta$ and $G_s=\Delta\Gamma$.
Now, from the Lemma 1, $\eta_k$, counted with their multiplicity, are eigenvalues of $G_s$. 

To prove the second part note that $|\alpha_k\rangle = \Gamma \theta_k$ is an eigenvector of $D$. In fact
$
D|\alpha_k\rangle = D\Gamma \theta_k = \Gamma G_s\theta_k = \eta_k\Gamma \theta_k=\eta_k|\alpha_k\rangle.
$
Now we have to show that if $|\eta_k\rangle$ is a non--zero eigenvector of $D$ it can be obtained by an appropriated $\theta_k$. Let $\theta_k=\Delta|\eta_k\rangle$, we have that 
$
G_s\theta_k=G_s\Delta|\eta_k\rangle=\Delta D|\eta_k\rangle = \eta_k\Delta|\eta_k\rangle=\eta_k\theta_k
$
then it is an eigenvector for $G_s$ and it cannot be zero, because otherwise
$
\Delta|\eta_k\rangle = 0\ \Rightarrow\ \Gamma\Delta|\eta_k\rangle=D|\eta_k\rangle=\eta_k|\eta_k\rangle=0
$
with contradiction. \q$\blacksquare$

\bigskip

\ev{Remark} We observe that if $\Gamma$ is full--rank, then SGM is full--rank from $\rank(\Gamma)=\rank(\Gamma^*\Gamma)=\rank(G_s)$.
In this case, $\rank(D)=\rank(G_s)$ follows from Lemma 1.
But it is not always true: in general we have $\im(D)\subseteq\im(\Gamma)$, then $\rank(D)\leq\rank(\Gamma)=\rank(G_s)$.
A simple example which leads to $\rank(D)<\rank(G_s)$ is provided by
\begin{equation}
\Gamma = \left[\!\begin{array}{c|c}
\begin{smallmatrix}
 1 & 0 \\
 0 & 1 \\
 0 & 0 \\
 0 & 0 
\end{smallmatrix}
&
\begin{smallmatrix}
 1 & 0 \\
 0 & 0 \\
 0 & 1 \\
 0 & 0  
\end{smallmatrix}
\end{array}\!\right]\quad\textrm{with } k_0=k_1=2. 
\end{equation}

Here $\Gamma$ is not full--rank matrix and $\rank(G_s)=3$ whereas $\rank(D)=2$.

%
%

%


\section{Check of the result with pure states}
 
In the case of rank 1 (pure states) the weighted density operators have the form
\begin{equation} 
  \rho'_0=q_0\,|a\rangle\langle a|\vq \rho'_1=q_1\,|b\rangle\langle b|
\end{equation}
where $|a\rangle$ and $|b\rangle$  are normalized pure states. 
 
\subsection{Helstrom's approach}
    
The decision operator $D$ is given by
\begin{equation}
  D= \rho'_1-\rho_0'=q_1\,|b\rangle\langle b| -q_0\,|a\rangle\langle a|
\end{equation}
and has rank 2 (if $|b\rangle\langle b|$ and $|a\rangle\langle a|$
are linearly independent). For the EID of $D$ we have to know the
components of the kets $|a\rangle$ and $|b\rangle$, namely
\begin{equation}
    |a\rangle=\left[a_0\ a_1\ \cdots\ a_{n-1}\right]^\intercal,\ 
    |b\rangle=\left[b_0\ b_1\ \cdots\ b_{n-1}\right]^\intercal\,.
  \label{eq:D18}
\end{equation}
Hence $D_{ij}=q_1 b_i^*b_j-q_0 a_i^*a_j$, $ i,j=0,1,\ldots,n-1$,
where the components are linked by the normalization conditions
$\langle a |a\rangle=\langle b |b\rangle=1$ and $q_0+q_1=1$.
But the explicit evaluation of the eigenvalue equation $\det(D-\eta I)$
is not immediate, since it involves a determinant of an
$n\times n$ matrix. So, another approach is followed \cite{Helstrom2}, based on the
fact that  $D$ has rank 2  and the eigenvectors $|\eta_i\rangle$  must belong to the
two dimensional subspace spanned by $D$, so they are a linear combination
of the states, namely $|\eta_i\rangle=A_i|a\rangle+B_i|b\rangle$.
Introducing this linear combination into the eigenvalue/eigenvector definition
$D|\eta_i\rangle=\eta_i|\eta_i\rangle$ and considering the normalization
condition, one gets $q_1(A_i\,X^*+B_i)|b\rangle-q_0(A_i+B_i\,X)|a\rangle
=\eta_i( A_i|a\rangle+B_i|b\rangle)$
where $X=\langle a|b\rangle$.
 Finally, considering that the states are linearly independent, so that their coefficients
 must be equal, we obtain the equations
\begin{equation}
  \begin{array}{l}
  q_1(A_i\,X^*+B_i)=\eta_i\,B_i \\
  q_0(A_i+B_i\,X)=\eta_i\,A_i
 \end{array} \vq i=1,2
\end{equation}
whose solutions give the  non--zero eigenvalues: 
$\eta_{\pm}=\frac12(q_1-q_0$ $\left.\mp\sqrt{1-4q_0q_1|X|^2}\right)$,
where $\eta_+>0$, $\eta_-<0$.
Hence, from (\ref{eq:DD12}) the correct decision probability is 
$P_{c}=q_0+\eta_+$,
that is 
\begin{equation}
 P_{c}=\fract12\left(1+\sqrt{1-4q_0q_1|X|^2}\right)
\label{eq:C8}
\end{equation}
which represents the so called {\it Helstom's bound}.

\subsection{Skew Gram matrix (SGM) approach}

Considering
the normalization, $\langle a|a \rangle=\langle b|b\rangle=1$, and
the notation $X=\langle a|b\rangle$,
 the skew Gram matrix is
\begin{equation}
 G_s=\left[\begin{array}{cc}-q_0 & -\sqrt{q_0q_1}\,X\\
\sqrt{q_0q_1}\,X^*&q_1\end{array}\right]\,.
\label{eq:P28}
\end{equation}
Then, equation (\ref{eq:D14}) becomes
\begin{equation}
\det\left[\begin{array}{cc}-q_0-\eta &-\sqrt{q_0q_1}X\\
	        \sqrt{q_0q_1}X^*&q_1-\eta \end{array}\right]=0
\label{eq:9a} 
\end{equation}
which has exaclty the  solutions $\eta_{\pm}$
  obtained above.

\ev{Comparison} We have seen that  Helstrom's approach is very
articulated, because it starts  with data expressed in terms of
 outer products,
whereas the final result is expressed in terms of inner products (the parameter
$X=\langle a|b\rangle$ in the case of pure states). The SGM approach
is straightforward because it
starts directly with  inner products, the same we find in
the final result. Note also that in the Helstrom's approach the original data
are redundant (states of size $n$), whereas the system performance are
completely determined by the geometry of inner products.

\section{An application to quantum state comparison}

A problem of particular interest is quantum state comparison,
where one wants to determine whether the states of a quantum system
are identical or not \cite{Bergou}. It can be formulated as a binary
problem of discrimination between a pure state $|a\rangle$ 
and a {\it uniformly} mixed state, namely
		  
\begin{equation}
   \rho_0=|a\rangle\langle a|\vq \rho_1=
  \frac1h\sum_{i=1}^h |b_i\rangle\langle b_i|
  \label{eq:H3}
\end{equation}
where the $h$ kets are supposed orthonormal, i. e. 
$\langle b_i|b_j\rangle=\delta_{ij}$.

We evaluate the probability of correct discrimination using the
SGM approach. The factors of the above density operators are
$
\gamma_0=\sqrt{q_0}\;|a\rangle$ and $ \gamma_1=\sqrt{q_1/h}\;[|b_1\rangle,
\ldots,|b_h\rangle]$
and the corresponding SGM is
\begin{equation}
 G_s=\left[\begin{array}{cc}
 -q_0 & -V \\ V^* & (q_1/h)I_h
\end{array}\right]
\end{equation}
where $V=\sqrt{q_0 q_1/h}\,[X_1,\ldots,X_h]$ with 
$X_i=\langle a|b_i\rangle$ and $I_h$ is the $h\times h$ identity matrix. 
The eigenvalue equation (\ref{eq:D14}) gives
\begin{equation}
\left(\frac{q_1}h-\eta\right)^{h-1}
\left[\left(\eta-\frac{q_1}h\right)(\eta+q_0)+\frac{q_0 q_1}h\left\|X\right\|^2\right]=0
\end{equation}
where $\left\|X\right\|^2=\sum_{i=1}^h|X_i|^2$.
Its $h+1$ solutions are
\begin{equation}
 \begin{split}
\eta_i & =\frac{q_1}h \vl i=1,\ldots,h-1\\
\eta_{\pm} & =\frac12 \left[\frac{q_1}h-q_0\pm\sqrt{\left(\frac{q_1}h+q_0\right)^2-\frac{4q_0q_1}h\left\|X\right\|^2}\right]
 \end{split}
\end{equation}
and are all positive, $\eta_-$ excepted. Hence, (\ref{eq:DD12}) gives
\begin{equation}
P_c=\frac12\left[1\!+\!q_1\!\frac{h\!-\!1}h\!+\!\sqrt{\left(\frac{q_1}h\!+\!q_0\right)^2\!-\!\frac{4q_0q_1}h\!\left\|X\right\|^2}\right].
\label{eq:S12c}
\end{equation}
The case of interest is when all the $h+1$ kets in (\ref{eq:H3}) have the same probability,
that is when $q_1=h/(h+1)$. In this case the error probability is
 $P_e=1/(h+1)\left(1-\sqrt{1-\left|X\right\|^2}\right)$.

Comparison with  \cite{Bergoubook}, where the direct
Helstrom approach is used, shows that the SGM approach is much more simpler.

\section{General case of  rank--2 density operators }

We develop the performance evaluation of the binary quantum system when the density
operators $\rho_0$ and $\rho_1$ have rank 2. A particular attention is paid
for the  reduction of the number of parameters
to get compact and  readable results.

\subsection{Helstrom's approach}
The weighted density operators have the form $\rho'_0=q_0\,(p_a\,|a\rangle\langle a|+p_c\,|c\rangle\langle c|)$, 
$\rho'_1=q_1\,(p_b\,|b\rangle\langle b|+p_d\,|d\rangle\langle d|)$
where $p_a$, $p_b$, $p_c=1-p_a$, and $p_d=1-p_b$ are the probabilities of the normalized states
$|a\rangle$, $|b\rangle$, $|c\rangle$, and $|d\rangle$,
respectively. In an $n$-dimensional Hilbert space the states have $n$
components: $a_i$, $b_i$, $c_i$, $d_i$, $i=0,1,\ldots,n-1$.
 Without loss of generality we can assume that the
states forming the  same density operator are orthogonal
\cite{Hughston},
that is
\begin{equation}
 \langle a|c\rangle=0\vq \langle b|d\rangle=0\,.
\label{eq:D4}
\end{equation}

With the above specifications the decision operator becomes
\begin{equation}
 \begin{split}
  D& =q_1\rho_1-q_0\rho_0\\
   & =q_1\left[p_b\,|b\rangle\langle b|\!+\!p_d\,|d\rangle\langle d|\right]-q_0\left[p_a\,|a\rangle\langle a|\!+\!p_c\,|c\rangle\langle c|\right] 
\end{split}
\label{eq:H13}
\end{equation}
and its  entries are
$D_{ij}=q_1[p_b\, b_ib_j^*+p_d\, d_id_j^*]-
        q_0[p_a\, a_ia_j^*+p_c \,c_ic_j^*]$.
Considering that $q_0+q_1=1$, $p_a+p_c=1$ and $p_b+p_d=1$,
the specification of the decision operator is given by
$q_0$, $p_a$, $p_b$, $|a\rangle$, $|b\rangle$, $|c\rangle$, $|d\rangle$,
where the kets are normalized and verify conditions (\ref{eq:D4}).

For the evaluation of the eigenvalues of $D$  we follow the procedure seen with
pure states. Now,
$D$ has rank 4  and the eigenvectors $|\eta_i\rangle$  must belong to the
four dimensional subspace spanned by $D$, so they are a linear combination
of the states, namely 
$|\eta_i\rangle=
  A_i|a\rangle+B_i|b\rangle+C_i|c\rangle+D_i|d\rangle
$.
Hence, introducing this linear combination into the eigenvalue/eigenvector definition
$D|\eta_i\rangle=\eta_i|\eta_i\rangle$ and considering the normalization
condition  and orthogonality (\ref{eq:D4}),  one gets
\begin{equation}
\begin{split}
   &A_i\,q_1\,p_a\,|a\rangle-A_i [\,q_0\,p_b X^*\,|b\rangle+\,p_d Y^*\,|d\rangle]+\\
  +&B_i\,q_1[\,p_aX\,|a\rangle+\,p_cW\,|c\rangle]-B_i\,q_0\,p_b\,|b\rangle+ \\
  +&C_i\,q_1\,p_c\,|c\rangle-C_i\,q_0[\,p_bW^*\,|b\rangle+\,p_dZ^*\,|d\rangle]+ \\
  +&D_i\,q_1[\,p_aY\,|a\rangle+\,p_cZ\,|c\rangle]-D_i\,q_0p_d\,|d\rangle \\
  =&\eta_i[A_i\,|a\rangle+B_i\,|b\rangle+C_i\,|c\rangle+D_i\,|d\rangle
\end{split}
\end{equation}

where we have introduced the inner products
\begin{equation}
 X=\langle a|b\rangle,\ Y= \langle a|d\rangle,\ W=\langle c|b\rangle,\ Z=\langle c|d\rangle\,.
 \label{eq:EE2}
\end{equation}

 Finally, considering that the 4 states are linearly independent, so that their coefficients
 must be equal, we obtain 4 equations in the unknowns $A_i,B_i,C_i,D_i$ and
 $\eta_i$. By solving with respect to $\eta_i$ we obtain an algebraic
 equation of degree 4
whose solutions give the  non zero eigenvalues.
The procedure is cumbersome and we do not write explicitly the passages,
since they are immediate from the SGM procedure.

\subsection{Skew Gram matrix  approach}

We have to find the factors of the density operators.
Considering the orthogonality condition (\ref{eq:D4}), they are given by
\begin{equation}
 \gamma_0=q_0\left[\sqrt{p_a}\,|a\rangle, \sqrt{p_c}\,|c\rangle\right],\ 
 \gamma_1=q_1\left[\sqrt{p_b}\,|b\rangle, \sqrt{p_d}\,|d\rangle\right]\,.
\label{eq:D6}
\end{equation}

Considering again orthogonality (\ref{eq:D4}) and  normalization, the 
     blocks of the SGM are given by
\begin{align}
\nonumber  G_{00}&=\gamma_0^*\gamma_0=q_0 \left[\begin{array}{cc}p_a&0\\ 0&\,p_c\end{array}\right], & G_{01}&=\gamma_0^*\gamma_1=\left[\begin{array}{cc}x&  y\\ w&z\end{array}\right],\\
  G_{11}&=\gamma_1^*\gamma_1= q_1 \left[\begin{array}{cc}p_b&0\cr 0&p_d\end{array}\right], &
  G_{10}&=\gamma_1^*\gamma_0=G_{01}^*, 
\label{eq:D12}
\end{align}
where
%
\begin{equation}
\begin{split}
  x&=\sqrt{q_0q_1 p_ap_b}X,\quad y=\sqrt{q_0q_1 p_ap_d}Y,\\
w&=\sqrt{q_0q_1 p_cp_b}W,\quad z=\sqrt{q_0q_1 p_cp_d}Z,
\end{split} 
\end{equation}
are weighted inner products.
We see that in general the SGM $G_s$ depends on the seven parameters $q_0$, $p_a$, $p_b$, $x$, $y$, $w$, $z$.
The eigenvalue equation  of $G_s$ is
\begin{equation}
  \eta^4+B\eta^3+C\eta^2+D\eta+E=0
 \label{eq:DA3}
\end{equation}
where
\begin{equation}
 \begin{split}
  E&=|xz|^2+|wy|^2-|x|^2 q s-|w|^2 q r-|y|^2 p s-|z|^2 p r+\\
  &\quad -x w^* z y^*-x^* w z^*y +p q r s\\
D&=|x|^2 (q-s)+|w|^2( q-r)+|y|^2(p-s)+\\
&\quad+|z|^2 (p-r)-p r q+p r s\\
C&=|x|^2\!+\!|w|^2\!+\!|y|^2\!+\!|z|^2\!+\!p q\!-\!p r\!-\!q r\!-\!p s\!-\!q s\!+\!r s\\
B&=p+q-r-s=q_0-q_1
 \end{split}
\end{equation}
%
with
\begin{equation}
   p=q_0p_a,\ q=q_0p_c,\ r=q_1p_b,\ s=q_1p_c\,.
   \label{eq:DA17}
\end{equation}

Equation (\ref{eq:DA3}) is a
 quartic equation, whose analytical solutions are known from the times
 of Cardano, but their expression are  very long.
When the states have equal prior probabilities we find $B=0$ and
  (\ref{eq:DA3}) becomes the {\it depressed quartic equation}
   for the absence  the third degree term. The depressed quartic equation
   played an important role in the
   history of mathematics.
However, we need the sum $\eta_1+\eta_2$ of the two positive solutions, which is more readable
than the individual solutions. We find
%
\begin{align}
\label{eq:DA19}
  &\eta_1+\eta_2 =\\
\nonumber&\sqrt{\frac{B^2}{4}\!-\!\frac{2 C}{3}\!+\!\frac{\sqrt[3]{R\!+\!\sqrt{R^2\!-\!4 S^3}}}
{3 \sqrt[3]{2}}\!+\!\frac{\sqrt[3]{2} S}{3 \sqrt[3]{R\!+\!\sqrt{R^2\!-\!4 S^3}}}}\!-\!\frac{B}{2}
\end{align}
where $ R=2 C^3-9 B D C-72 E C +27 D^2+27 B^2 E$ and $S=C^2-3 B D+12 E$.

\subsection{Interpretation of parameters and numerical example}

We can take as reference the case of {\it pure states} $|a\rangle$ and
$|b\rangle$ and their inner product $X=\langle a|b\rangle$, which
determines the error probability through $|X|^2$ according to (\ref{eq:C8}).
In the ideal case
the states are orthogonal, $X=0$, and give $P_c=1$. When $|X|>0$
we find  $P_c<1$ and the degradation is to ascribe to the
presence of the {\it shot noise} (in the classical interpretation).

Now, it is difficult to choose the data listed in (\ref{eq:DA17})
to carry out  a numerical example, owing to the several conditions therein,
which ultimately assured
 that the correspondent Gram matrix is PSD. To be sure that
 the data are feasible, we have chosen the density operators from Glauber's
 theory on
 {\it coherent states}, modeling  the monochromatic
 electromagnetic radiations produced by a laser. In these theory
a density operator $\rho(\alpha)$, represented by a matrix of infinite
dimension, depends only on two parameters $\alpha$ and $\C(N)$, with
 $N_\alpha=|\alpha|^2$ giving the average number of {\it signal photons},
and $\C(N)$ the   average number of {\it thermal noise photons}. For $\C(N)=0$
(absence of thermal noise)
we find rank 1, that is pure states. Thus, to find rank 2 we have 
to consider a small amount of thermal noise. Note that for practical calculations,
we need a finite $n$--dimensional approximation,   
of the  infinite dimensional representation. As  
discussed in detail  in \cite{CarPier}, to get a moderately small value of the
size $n$ we have to choose a small value of $N_\alpha=|\alpha|^2$
and  a rank 2 is assured with a very small value of $\C(N)$.  In practice,
a good choice  for a non symmetric  case may be $n=10$, $q_0=0.4$, $\alpha_0=-1.2247 \arr N_{\alpha_0}=1.5$,
$\alpha_1 =1.3038\arr N_{\alpha_1}=1.7$, $\C(N)=0.05$.

With these parameters we have evaluated the $10\times10$ density operators 
$\rho_0=\rho(\alpha_0)$, $\rho_1=\rho(\alpha_1)$, and the $10\times2$
factors $\gamma_0$, $\gamma_1$. For reason of space we omit
the explicit numerical matrices and we write directly the
 blocks of the SGM, which result
\begin{align}
  \nonumber G_{00}&=\left[\begin{array}{cc}0.381 & 0 \\ 0& 0.018\end{array}\right], & 
  G_{01}&=\left[\begin{array}{cc}0.019 & 0.011 \\0.011 & 0.005\end{array}\right], \\
  G_{11}&=\left[\begin{array}{cc}0.571 & 0 \\0 & 0.027\end{array}\right], &  
  G_{10}&=G_{01}^*\,.
\label{eq:S13A}
\end{align}
%

From (\ref{eq:S13A}) we obtain (see (\ref{eq:D12}))
the  probabilities
$
p_a=0.95240$, $ p_c=0.04534$, $ p_b=0.95243$, $ p_d=0.04533
 $
and the  inner products
$
X=0.04089$, $ Y=0.10346$, $ W=0.10345$, $ Z=0.22043
 $.
We have evaluated the 10 eigenvalues of $D$, which result 
$\{-0.380307$, $-0.0174054$, $0$, $0$, $0$, $0$, $0$, $0$, $0.0263829$, $0.570885\}$
and give
$
P_c=0.997268\vl P_e=1-P_c=0.00273197
 $.
We have checked that the 4  eigenvalues of $G_s$ are exactly the non zero eigenvalues 
of $D$, in agreement
with the theory.
The coefficients of the eigenvalue equation (\ref{eq:DA3}) are
$ E=0.000105191$, $D= 0.00183749$, $C=-0.216064$, $B=-0.198617$
and the solutions are $\{-0.380307$, $-0.0183563$, $0.0263943$, $0.570885\}$.
 To get the sum of the positive solutions (\ref{eq:DA19}) from the formulas, we have
 evaluated
 $R=-0.0190434$, $S=0.0490408$. Then
$
\eta_1+\eta_2=0.59728
 $,
 which gives again tha above probabilities.

The error probability with pure states, obtained with the same inner product $X=0.04089$, is $P_e=0.000401349,$ that is one order of magnitude better than with rank 2.  
The degradation with  mixed states is due to the presence of thermal noise.

\section{Rank 2 density operators with symmetry}

An $m$--ary state constellation $\{\gamma_0,\gamma_1,\ldots,\gamma_{m-1}\}$ 
 exhibits the  {\co geometrically uniform symmetry} (GUS), see \cite{Elder}, if the states
 are related as  
$\gamma_i=S^i\gamma_0$, where $S$ is
 a unitary operator %
$S$, such that $S^m=I_\C(H)$ is the identity operator in the Hilbert space $\C(H)$. With the GUS equal prior probability are assumed, 
 that is $q_i=1/m$.

 Helstrom's  approach is not very much simplified by the GUS, so we pass directly
to the alternative equivalent approach.

\subsection{SGM approach with GUS}

We investigate the simplifications on the Gram matrices due to the GUS and  
 equal prior probabilities. We know that, in the presence of GUS,
the Gram matrix becomes {\it block--circulant}, that is its blocks $G_{ij}$
depends only on the index differences $j-i$ \cite{CarPier}. In the binary case
the block--circulant conditions are
$
G_{11}=G_{00}$, $ G_{01}=G_{10}=G_{01}^*$.
Inspection on (\ref{eq:D12}) shows that this condition 
leads to the following simplifications:
 1)  $p_b=p_a$ and $p_d=p_c$,
 2) the inner product $X=\langle a|b\rangle\vl Z=\langle c|d\rangle$
are real,
 3) the inner products $Y$ and $W$ are conjugate. 
Hence, the leading blocks take the form
\begin{equation}
 G_{00}\!=\!\frac12\!\left[\!\begin{array}{cc}
p_a & 0 \\
 0 & p_c
\end{array}\!\right]\!,
G_{01}\!=\!\frac12\!\left[\!\begin{array}{cc}
p_a\,X & \sqrt{p_ap_c}\,Y \\
\sqrt{p_ap_c}\,Y^* & p_c\,Z
\end{array}\!\right]
\label{eq:D17}
\end{equation}
where $p_a+p_c=1$. In conclusion
with the GUS the SGM depends only on the four parameters
$
p_a, X, Y, Z
$.

 The  eigenvalue equation of $G_s$, given by (\ref{eq:9a}), written in a convenient form, is
\begin{equation}
 \eta^4-\fract14\, H\,\eta^2+\fract1{16}\,L=0
\label{eq:DD8} 
\end{equation}
where
\begin{align}
 H & = p_a^2(1-X^2)+p_c^2(1-Z^2)-2p_ap_c|Y|^2\cr
 L & = (p_ap_c)^2\left[|Y|^4-2(1+XZ)|Y|^2+(1-X^2)(1-Z^2)\right]\cr
  & = (p_ap_c)^2\left[(|Y|^2-(1+XZ))^2-(X+Z)^2\right]\,.
  \label{eq:DD9}
\end{align}
 Hence, the quartic equation (\ref{eq:DA3}) degenerates into a bi--quadratic 
  equation, whose  solutions are straightforward. In particular, the
 two positive solutions are given by
$
\eta_{1,2}=\fract12\sqrt{\fract12(H\pm\sqrt{H^2-4 L})}\,,
$
hence
\begin{align}
 P_c&= \fract{1}{2} +\eta_1+\eta_2 \cr
    &=\fract{1}{2}\!+\!\fract12\!\sqrt{\fract12(H+\sqrt{H^2-4 L})}\!+\!\fract12\!\sqrt{\fract12(H-\sqrt{H^2-4 L})}\cr
     &= \fract{1}{2}+\fract{1}{2}\sqrt{H +2\sqrt{L}}.
\end{align}
The explicit result is
\begin{multline}
\label{eq:T21}
P_c=\fract12\!+\!\fract12\left(p_a^2(1\!-\!X^2)\!+\!p_c^2(1\!-\!Z^2)\!-\!2p_ap_c|Y|^2+\right. \\
\left.\!+\!2 p_ap_c\sqrt{[|Y|^2\!-\!(1\!+\!XZ)]^2\!-\!(X\!+\!Z)^2}\right)^{1/2}.
\end{multline}

\ev{Check with pure states} The case of pure states is obtained by letting
$p_a=1$ and then $p_c=0$.  In agreement with (\ref{eq:C8}), from  (\ref{eq:T21})  we get
\begin{equation}
 P_c=\fract12+\fract12\sqrt{1-X^2}\,.
\end{equation}

\ev{Case of orthogonality} If $Y=\langle a|d\rangle=0$,
also $W=\langle c|b\rangle=0$. This leads to a simplification in (\ref{eq:T21}),
 namely
\begin{equation}
  P_c =\fract12+\fract12\left(p_a\sqrt{1-X^2}+p_c\sqrt{1-Z^2}\right)\,.
\end{equation}

  \subsection{Numerical example}

A  choice  for a  symmetric case may be  $n=10$, $q_0=0.5$, $\alpha_0=-1.26491 \arr N_{\alpha_0}=1.6$,
$\alpha_1 =1.26491\arr N_{\alpha_1}=1.6$, $\C(N)=0.05$ then, the leading blocks of the Gram matrices are
\begin{equation}
 \begin{split}
  G_{00}&=\gamma_0^*\gamma_0=\left[\begin{array}{cc}0.476206 & 0 \cr  0 & 0.0226692\end{array}\right], \\
  G_{01}&=\gamma_0^*\gamma_1=\left[\begin{array}{cc}0.019409 & -0.0107206 \cr  -0.0107206 & 0.00498695\end{array}\right]\,.
 \end{split}
\label{eq:S13B}
\end{equation}

From (\ref{eq:S13B}) we obtain (see (\ref{eq:D12}))
the  probabilities:
$
p_a=0.95241$, $ p_c=0.04534$, $ p_b=0.95241$, $ p_d=0.04534
 $
and the  inner products
$
X=0.04076$, $ Y=-0.10318$, $ W=-0.10318$, $ Z=0.21999
 $.
The eigenvalues of $D$ are $\{-0.47558$, $-0.0218743$, $0$, $0$, $0$, $0$, $0$, $0$, $0.0218743$, $0.47558\}$
and give
$
P_c=0.997268$ and $P_e=0.00273197
 $.
The  eigenvalues of $G_s$ are exactly the non zero eigenvalues of $D$,
 in agreement with the theory.
The coefficients of the eigenvalue equation (\ref{eq:DD8}) are
$-\fract1{16}\,L= 0.000108222$, $\fract14H=-0.226655$
 and the solutions $\{-0.47558$, $-0.0218743$, $0.0218743$, $0.47558\}$.

We now check that the same results are obtained from the analytical formulas.
The evaluation of (\ref{eq:DD9}) gives $
H=0.907538
 $ and $L=0.0017714$. Hence, from (\ref{eq:T21}) we obtain
$P_c=0.997924$, in agreement with the above evaluation.

The case $Y=0$ gives $
H=0.907538 $,
$L=0.0017714$, and
$P_c=0.997924$, $P_e=0.00207581 $.

The  error probability with pure states (with
  the same inner product $X=0.04076$) is $P_e=0.000415467$, that is one order
  of magnitude better than with with with rank 2.  The degradation is due to the
  presence of thermal noise.

\vspace{3mm}

\subsection{Limit to closed--form results} 
As established by Evarist Galois
two hundred years ago, the solution of algebraic equations can be written
explicitly up to the fourth order. We have seen that
in the general case of the binary quantum detection the case of rank 2+2
leads just to a quartic equation and so it is the limit. In the presence
of GUS  the quartic  equation is essentially reduced to the second order.
In principle, with the GUS it would be possible to solve the case of
rank 3+3, where the six order equation can be reduced to a cubic,
and also the case of rank 4+4, where the eight order equation can be reduced 
to a quartic. No more closed form is possible, unless in particular cases.

\section{Conclusions}
  
  The Helstrom theory on binary detection is the only general explicit result
  available in quantum optimization, but its  translation into
  formulas
  is really possible only with very small dimensions of the density operators.
   On the other
  hand, the final results depends only on the inner products of quantum states,
  which compress the information contained in the density operators.
  The skew Gram matrix approach, which gives exactly the same results,
  starts just from the inner products, with a dramatic simplification
  of the algebra involved, and permits to establish closed--form
  results not available elsewhere. Furthermore, the skew Gram matrix approach
   clearly establishes a
  fundamental truth, at least for binary detection: the inner products,
  collected in the Gram matrix, give the necessary and sufficient information
  to achieve the optimal detection.

\end{document}